\documentclass[twocolumn,aps,showpacs,epsfig]{revtex4}%
\usepackage{amsfonts}
\usepackage{amsmath}
\usepackage{amssymb}
\usepackage{graphicx}%
\begin{document}
\title{On the residual resistivity near a two dimensional \\
metamagnetic quantum critical point}
\author{Yong Baek Kim$^{(a)}$ and A. J. Millis$^{(b)}$}

\affiliation{$^{(a)}$Department of Physics, University of Toronto, Toronto, 
Ontario M5S 1A7, Canada\\
$^{(b)}$ Department of Physics, Columbia University, 538 W. 120th St.,
New York, New York 10027}
\date{\today}

\begin{abstract}
The behavior of the residual (impurity-dominated) resistivity is computed for
a material near a two dimensional quantum critical point characterized by a
divergent $q=0$ susceptibility. A singular renormalization of the amplitude
for back-scattering of an electron off of a single impurity is found. When the
correlation length of the quantum critical point exceeds the mean free path,
the singular renormalization is found to convert the familiar
`Altshuler-Aronov' logarithmic correction to the conductivity into a 
squared-logarithmic form. Impurities can induce unconventional Friedel 
oscillations, which may be observable in scanning tunnelling microscope experiments. 
Possible connections to the metamagnetic quantum critical end point recently proposed
for the material $Sr_{3}Ru_{2}O_{7}$ are discussed.

\end{abstract}
\maketitle

\section{Introduction}

The interplay of quantum mechanics, disorder and reduced dimensionality is a
central question in condensed matter physics. The issue becomes particularly
interesting for materials near a \textit{quantum} critical point (i.e. one
produced at temperature $T=0$ by variation of a control parameter such as
pressure or chemical composition). Near such a critical point, quantum
fluctuations may become particularly strong, and may interact with randomness
in an important way. In this paper, we consider the heretofore little studied
question of the effect of quantum critical fluctuations on the residual
resistivity of a metal. We consider two dimensional systems in which the
criticality involves long wavelength fluctuations and find that the critical
fluctuations lead to a singular, and possibly divergent, renormalization of
the amplitude for an electron to backscatter off of an isolated impurity atom.
For a system with a nonvanishing density of impurities, 
we find that this physics leads to a
strengthening of the 
`Altshuler-Aronov' \cite{Altshuler80} correction to the conductivity
from $ln(1/T)$ to $ln^2(1/T)$.

Renormalization of some electron-impurity vertices is expected in materials
near density wave transitions; for example, near a charge density wave
transition, an impurity will produce density fluctuations whose component at
the ordering wavevector will diverge as the transition is approached (for an
experimental demonstration, see \cite{Wertering98}). However, we show here
that long-wavelength spin fluctuations can drastically increase
backscattering by spinless impurities. Our work is related to a previous
analysis, by Altshuler, Ioffe. Larkin, and one of us, of the staggered
susceptibility of a model of electrons interacting with gauge fluctuations
\cite{Altshuler95}.

Our work is motivated by recent experiments on the bilayer ruthenate,
$Sr_{3}Ru_{2}O_{7}$ , which apparently can at ambient pressure be
tuned through a quantum critical point by variation of applied magnetic field
\cite{Perry01,Grigera01,Chiao02}. $Sr_{3}Ru_{2}O_{7}$ has a layered crystal
structure and highly anisotropic conductivity properties charactered by a very
low in-plane (`ab') resistivity ($\rho_{ab}\sim5\mu\Omega-cm$
as $T\rightarrow0$) and a much higher out of plane (`c-axis') resistivity
($\rho_{c}\sim10^{3}\mu\Omega-cm$ as $T\rightarrow0$) \cite{Ikeda00}, so its
electronic properties are to a good approximation two dimensional. At ambient
pressure and no applied magnetic field the material is paramagnetic, but with
a magnetic susceptibility $\chi$ which is strongly $T-$dependent (for
$T\gtrapprox10K$) and at low $T$ strongly enhanced above the band theory value
\cite{Ikeda00}. As applied magnetic field $H$ is increased at low $T$ ($T<2K$)
resistivity measurements indicate
that the material passes near to a critical point \cite{Perry01,Grigera01}.
In the vicinity of this critical point 
the differential
magnetic susceptibility $\chi_{\mathrm{diff}}=dM/dH$ becomes large
and the residual resistivity $\lim_{T\rightarrow
0}\rho(T)$ exhibits a pronounced cusp \cite{Grigera01}.

The natural interpretation is that the
transition observed in $Sr_{3}Ru_{2}O_{7}$ is \textit{metamagnetic} in
nature. Metamagnetic materials exhibit a phase diagram in the field
($H$)-temperature ($T$) plane characterized by a line of first order
transitions across which the magnetization jumps. The first order line ends in
a critical point at temperature $T_{M}$ and field $H_{M}$. In $Sr_{3}%
Ru_{2}O_{7}$ parameters are apparently such that at ambient pressure $T_{M}$
is very near to $T=0$ (indeed for one field orientation it can be made to
vanish exactly ) so that one has a \textit{quantum critical end point}
\cite{Perry01,Grigera01,Millis02a,Chiao02}. A renormalization group theory of
quantum critical end-points and their application to $Sr_{3}Ru_{2}O_{7}$ was
given by \cite{Millis02a}; in this paper we extend their theory to include
electron-impurity coupling and use it to calculate the residual resistivity.
In addition we make a few remarks about the applicability of our calculations
to more general classes of critical points.

The rest of this paper is organized as follows. Section II presents the
critical theory, section III gives the analysis of the renormalization of the
Born approximation amplitude for an electron to scatter off of an isolated
impurity, section IV considers the single impurity problem beyond the Born
approximation and applies the results to $Sr_{3}Ru_{2}O_{7}$, section V treats
the extension to a non-vanishing density of impurities and section VI\ is a
conclusion and discussion of implications for other experiments.

\section{Metamagnetic Quantum Criticality}

Ref \cite{Millis02a} employed the standard approach of Hertz \cite{Hertz76}
and Millis \cite{Millis93} which involves integrating out electronic degrees
of freedom to obtain an effective field theory with overdamped bosonic
excitations, which is then analysed. This approach is presently the subject
of debate \cite{Coleman01}. On the experimental side, the theory is apparently
inconsistent with data obtained on a variety of materials with
antiferromagnetic critical points, and on the theoretical side the validity of
integrating out gapless electronic degrees has been questioned
\cite{Belitz98,Chitov01,Abanov01}, both two dimensional antiferromagnets and
for ferromagnetic transitions involving an order parameter with continuous
symmetry and time reversal invariance in the disordered phase. However, the
assumptions made in \cite{Belitz98,Chitov01} do not apply to metamagnetic
transitions, which involve an Ising order parameter and an explicitly broken
time reversal symmetry, nor do the Fermi surface nesting complications of
antiferromagnetic transitions \cite{Abanov01} apply here, because the order
parameter is centered at $q=0$.

The order parameter for a metamagnetic critical point occurring at a field
$H_{M}$ is the difference of the local magnetization density $M(\mathbf{x}%
,\tau)$ (which we write here in space and imaginary time) from the average
value $M^{\ast}$ produced by the field $H_{M}$ at $T=0$. ($M^{\ast}%
\approx.5\mu_{B}$ for $Sr_{3}Ru_{2}O_{7}$.) It is convenient to measure the
field and magnetization with respect to $H_{M}$, $M^{\ast}$, and to normalize
the field to an energy scale $E_{0}$ (discussed below) and the magnetization
fluctuations per $Ru$ via
\begin{align}
\phi(\mathbf{x},\tau)  &  =\frac{M(\mathbf{x},\tau)-M^{\ast}}{M_{sat}%
},\label{phidef}\\
h  &  =\frac{(H-H_{M})M_{sat}}{E_{0}}, \label{hdef}%
\end{align}
where $M_{sat}$ is the high field saturation magnetization. In a clean
(non-disordered) system the action describing the fluctuations of $\phi$ was
argued \cite{Millis02a} to be ($\beta= 1/T$)%
\begin{equation}
S=\frac{E_{0}}{2}\int\frac{d^{2}x}{a^{2}}\int_{0}^{\beta}d\tau\left[  2
h\phi+\xi_{0}^{2}\left(  \nabla\phi\right)  ^{2}+\frac{1}{2}\phi^{4}\right]
+S_{dyn} \label{S}%
\end{equation}
with $\phi(\mathbf{q},\nu)=E_{0}a^{-2}\int d^{2}x\int_{0}^{\beta}d\tau
e^{i\mathbf{q\cdot x}-i\nu\tau}\phi(\mathbf{x},\tau)$ and, in a clean system,%
\begin{equation}
S_{dyn}=\frac{a^{2}}{2E_{0}}\int\frac{d^{2}q}{\left(  2\pi\right)  ^{2}}%
T\sum_{\nu_{n}}\frac{\left\vert \nu_{n}\right\vert }{vq}\left\vert
\phi(\mathbf{q},i\nu_{n})\right\vert ^{2} \ , \label{Sdyn}%
\end{equation}
where $\nu_{n} = 2 \pi n T$. For $Sr_{3}Ru_{2}O_{7}$, $E_{0}\approx7000K$
\cite{Millis02a} ; a value for $\xi_{0}$ has not been established. A natural
guess would be something of the order of the in-plane lattice constant
$a\approx4\mathring{A}$. However in many ferromagnetic materials band theory
suggests that the momentum dependence of the magnetic polarizability is very
weak, suggesting $\xi_{0} < a$; while if the considerations of \cite{Belitz98}
are relevant a considerably larger value might be appropriate. A value for $v$
has also not yet been established; the natural expectation is that it is of
order the planar-band Fermi velocity $v_{F}\approx1eV-\mathring{A}$. The
theory is above its upper critical dimension so the nonlinear term is a
`dangerously' irrelevant operator and may be treated by standard means
\cite{Millis93,Millis02a}.

We have written a two dimensional theory. At some scale a crossover to three
dimensional behavior will occur, but present neutron experiments have so far
been unable to observe any correlations along the c-axis \cite{Capogna02}.
Because the transition takes place in a magnetic field, the fluctuating field
has Ising symmetry and time reversal invariance is explicitly broken. Neither
`precession' terms in the dynamics nor the anomalous $\left\vert q\right\vert
$ momentum dependence proposed in \cite{Belitz98} will occur.

The action given in Eq \ref{S} describes the critical fluctuations. In the
absence of critical fluctuations we take the action for electrons to be%
\begin{equation}
S_{el}=\frac{a^{2}}{E_{0}^{2}} T\sum_{\omega_{n}}\int{\frac{d^{2}p}{(2\pi
)^{2}}}\psi_{\alpha}^{\dagger}(\mathbf{p},i\omega_{n})(-i\omega_{n}%
+\xi_{\mathbf{p}}) \psi_{\alpha}(\mathbf{p},i\omega_{n}), \label{Sel}%
\end{equation}
where $\psi_{\alpha}$ represents the electrons with the ``spin'' index
$\alpha$, $\xi_{\mathbf{p}}$ is the electron dispersion and $\omega_{n} =
(2n+1) \pi T$. The coupling between the electrons and the critical
fluctuations can be written as
\begin{equation}
S_{\phi-\psi}= g \int\frac{d^{2}x}{a^{2}} \int E_{0} d\tau\ \psi_{\alpha
}\sigma_{\alpha\beta}^{z}\psi_{\beta}\phi\ , \label{coupling}%
\end{equation}
where $\sigma^{z}$ is the $z$-component of the Pauli matrix.

To obtain the propagator $D$ describing the metamagnetic fluctuations we
expand the action about the mean field value $\phi_{0}$ given by $\partial S
/\partial\phi=0$ and read off the quadratic term, obtaining
\begin{equation}
D(\mathbf{q},i\nu_{n})=\frac{1}{\frac{\left\vert \nu_{n}\right\vert }{vq}%
+\xi_{0}^{2}q^{2}+h^{2/3}} \label{ddef}%
\end{equation}
The coupling constant $g$ may be determined because the damping term in
$S_{dyn}$ arises from processes in which a critical fluctuation decays into
one particle-hole pair \cite{Hertz76,Millis93} and is
\begin{equation}
g^{2}=\frac{4 \pi^{2}}{a^{2} E_{0}}\frac{v_{F}^{2}}{vS_{F}} \label{gdef}%
\end{equation}
with $S_{F}$ the length of the Fermi surface in two dimensions ($S_{F}=2\pi
p_{F}$ for a circular Fermi surface).

\begin{figure}[h]
\includegraphics[height=2.1cm,width=7cm,angle=0]{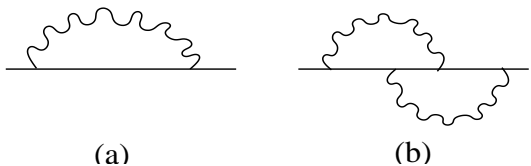}\caption{Diagrams
representing (a) the lowest order self-energy (b) an example of higher order
corrections. The wavy line represents the bosonic fluctuations.}%
\label{figure:1}%
\end{figure}

Using these definitions we find that the one-loop self energy at $T=0$, shown
in Fig 1a, is%
\begin{equation}
\Sigma_{1}(\mathbf{p},i\omega)=\frac{2 i v_{F}}{S_{F}\xi^{2}_{0}}\int^{\infty
}_{0} dx \ x \ \mathrm{ln} \left(  \frac{\frac{\omega\xi_{0}}{v} + x^{3} +
h^{2/3}x} {x^{3} + h^{2/3}x} \right)  . \label{sigmaoneloop}%
\end{equation}
Notice that as $\omega\rightarrow0$ at fixed $h\neq0$, $\Sigma_{1}%
(\mathbf{p},i\omega)=\frac{2v_{F}}{vS_{F}\xi_{0}}\frac{i\omega}{h^{1/3}}$
while as $h\rightarrow0$ at fixed $\omega$ we have $\Sigma_{1}(\mathbf{p}%
,i\omega)=\frac{isgn(\omega)\left\vert \omega\right\vert ^{2/3} 2 \pi\left(
v/\xi_{0}\right)  ^{1/3}}{\sqrt{3} S_{F}\xi_{0}(v/v_{F})}$.

In spatial dimension $d>2$ the fact that the theory is above the upper
critical dimension guarantees that a one-loop (Migdal-like) approximation
yields an asymptotically exact approximation to the self energy, but in $d=2$
this is not the case \cite{Altshuler94,Ioffe98,Kim94}. While higher
corrections such as those shown in Fig 1b do not change the powers, they do
introduce a dependence on momentum $\delta p=p-p_{F}$ and induce a dependence
on $\frac{\omega}{\omega^{\ast}(h)}$ with $\omega^{\ast}(h)=vh/\xi_{0}$. The
scaling function for the self-energy has only been computed in large $N$ and
small $N$ expansions where $N$ is the number of order parameter components.
Unfortunately, in the metamagnetic problem of interest here $N=1$.

\section{Electron-Impurity Vertex: Born Approximation}

\begin{figure}[h]
\includegraphics[height=5cm,width=6cm,angle=0]{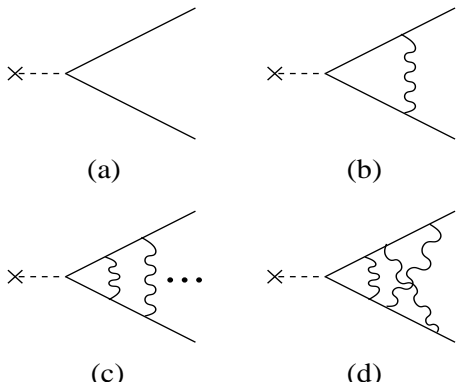}\caption{Diagrams
corresponding to the renormalization of the impurity scattering amplitude. (a)
the bare vertex (b) the lowest order correction (c) the sum of the ladder
diagrams (d) an example of higher order corrections. }%
\label{figure:2}%
\end{figure}

We now consider the effect of critical fluctuations upon the amplitude,
$A_{\mathbf{p}_{1},\mathbf{p}_{2}}$ for an electron initially in a state of
momentum $\mathbf{p}_{1}$ to scatter off of an isolated impurity atom into a
state of momentum $\mathbf{p}_{2}.$ The basic electron-impurity vertex is
shown in Fig 2a; we assume that this is structureless; corresponding to
scattering amplitude independent of $\mathbf{p}_{1,2}$. The first correction
due to critical fluctuations is shown in Fig 2b, and is found to be divergent
when $\mathbf{p}_{1,2}$ are on the Fermi surface and are such that the Fermi
surface tangents at these two points are parallel. For a circular Fermi
surface this condition corresponds to $\mathbf{p}_{1}+\mathbf{p}_{2}=0$ and
$\left\vert \mathbf{p}_{1}\right\vert =p_{F}.$ In this case, choosing the $x$
coordinate to be parallel to $\mathbf{p}_{1}$ and $\mathbf{p}_{1,2}$ to be on
the Fermi surface we have $\varepsilon_{\mathbf{p}_{1}+\mathbf{q}}%
=v_{F}\left(  q_{x}+q_{y}^{2}/2q_{0}\right)  $ and $\varepsilon_{\mathbf{p}%
_{2}+\mathbf{q}}=v_{F}\left(  -q_{x}+q_{y}^{2}/2q_{0}\right)  $ with $q_{0}$ a
quantity of the order of $p_{F}$ parametrizing the curvature of the Fermi
surface. If $\mathbf{p}_{2}=-\mathbf{p}_{1}+\mathbf{p}$ with $|\mathbf{p}%
|\ll|\mathbf{p}_{1}|$, we find that the the correction shown in Fig 2b is%
\begin{equation}
A_{1}(\mathbf{p},h)=A_{0}I\left(  \frac{2\pi q_{0}}{S_{F}}\right)  \ln\left(
\frac{1}{\max[h^{2/3},(p_{y}\xi_{0})^{2}]}\right)  \label{A2b}%
\end{equation}
with $I(b)=\frac{2b}{\pi}\int_{0}^{\infty}dy\ \frac{(\frac{2}{\sqrt{3}}%
b)y}{(1+y^{3})[(\frac{2}{\sqrt{3}}b)^{2}+y^{4}]}$ and $A_{0}$ is the bare
scattering amplitude; $I(1)\approx0.23$; $I(10)\approx0.5$ and $I(b)$ is an
increasing function of $b$. Notice that if $\pi-\theta$ is the angle between
$\mathbf{p}_{1}$ and $\mathbf{p}_{2}$, then $p_{y}\xi_{0}\propto\theta$.

A very similar result was obtained in the context of the $U(1)$ gauge theory
of the `RVB' regime of the two-dimensional $t-J$ model \cite{Altshuler94},
where the $2p_{F}$ spin susceptibility was considered. The $U(1)$ gauge theory
possesses an interaction (mediated by an internal gauge field) with a very
similar mathematical structure to our interaction $D ({\bf q},i\nu_n)$ 
(Eq.\ref{ddef}), except
that in the $U(1)$ problem gauge invariance dictates that in the $RVB$ regime
the mass ($h$ in our notations) vanishes. Further, in the vertex computation
two additional minus signs occur (but compensate each other), one from the
fact that gauge interaction involves currents which are oppositely directed at
momentum transfer $2p_{F}$ and one from the transverse nature of the gauge interaction.

Higher order diagrams such as those shown in Fig. 2c may be evaluated
similarly; we find that the leading behavior of the $n^{th}$ order term is
\begin{equation}
A_{n}=A_{0}\frac{1}{n!}\left[  I\left(  \frac{2\pi q_{0}}{S_{F}}\right)
\ln\left(  \frac{1}{\max[h^{2/3},\theta^{2}]}\right)  \right]  ^{n} \label{An}%
\end{equation}
so that finally we obtain%

\begin{equation}
A(\theta,h)\sim A_{0}\left(  \max[h^{2/3},\theta^{2}]\right)  ^{-\psi
}\label{A}%
\end{equation}
with $\psi=I\left(  \frac{S_{F}}{2\pi q_{0}}\right)  $ at one loop order.

We should further consider the effect of higher order diagrams, such as those
shown in Fig 2d. These do not change the basic power counting, and may
therefore be expected not to alter the basic power law we have found; they
will however change the numerical value of the exponent. According to the
result of small $N$ expansion \cite{Altshuler94}, the effect of `crossed
diagrams' is to increase the exponent. We conclude that the Born-approximation
amplitude for an electron to back-scatter off of an impurity suffers a
singular renormalization near criticality.

\section{Beyond Born Approximation; Resistivity}

The previous section presented the renormalization of the leading-order term
in the electron-impurity vertex. In this section we consider effects arising
when the Born approximation is not justified, either because the
initial scattering amplitude is not small or because the renormalizations
increase an initially weak interaction beyond the regime of validity of the
Born approximation. \ Corrections to the Born approximation result for the
electron self energy  correspond to multiple scattering of the electron off of
the same impurity, and are represented diagrammatically in Fig. 3.
Within the
`non-crossing' approximation used in previous sections we find that the
leading renormalization near criticality comes from the correcting each
impurity vertex individually. The series for the self energy may be summed by
defining the $T$-matrix which for incident electron energies very close to the
fermi surface becomes
\begin{widetext}
\begin{equation}
T(\theta-\theta^{\prime},h)=A(\theta-\theta^{\prime},h)
- i\int\frac{d\theta_{1}}{2\pi} N(\theta_{1})
A(\theta-\theta_1,h)T(\theta_{1}-\theta^{\prime},h)
\label{T}%
\end{equation}
\end{widetext}
with $N_{0}=\int\frac{d^{2}k}{\left(  2\pi\right)  ^{2}}\delta(\varepsilon
_{p})$ the single-spin fermi surface density of states. Assuming for
simplicity a circular fermi surface with density of state $N_{0}$ we may solve
the equation by resolving $T$ and $A$ into their angular components $T_{m}=\int
\frac{d\theta}{2\pi}T(\theta,h)e^{im\theta}$ and 
$A_{m}=\int \frac{d\theta}{2\pi}A(\theta,h)e^{im\theta}$ so that
\begin{equation}
T_{m}=\frac{A_{m}}{1+iA_{m}N_{0}}\label{Tm} \ .
\end{equation}
We note that the sign of $T_{m}$ and therefore the sign of the angular
momentum-$m$ channel phase shift alternates with $m$, so that the Friedel sum
rule is straightforwardly satisfied. 

\begin{figure}[h]
\includegraphics[height=1.7cm,width=8cm,angle=0]{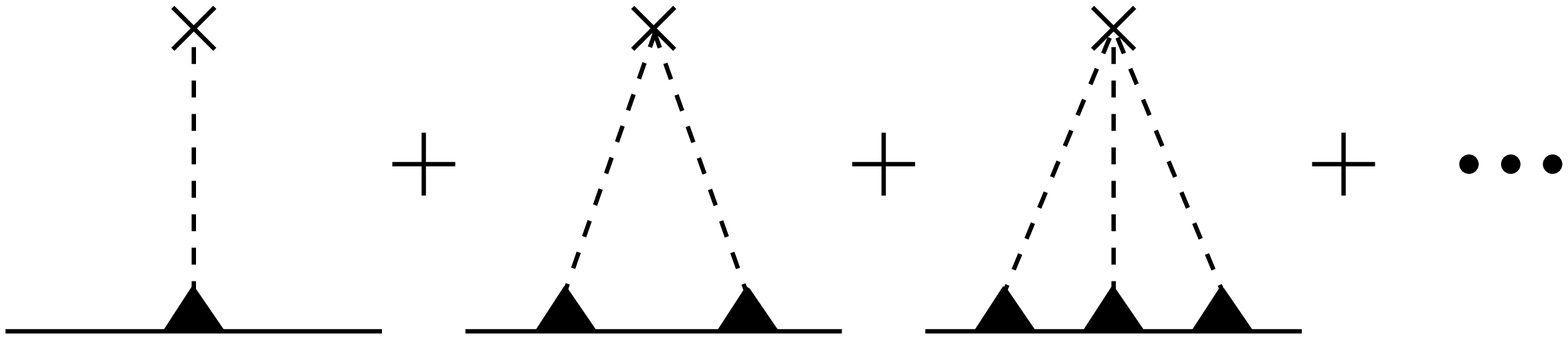}
\caption{Diagrams representing multiple scattering of electrons off of an impurity. 
The shaded triangle represents the renormalization of the Born scattering 
amplitude given by Fig.2c.}
\end{figure}
 
The final result depends on the parameter $g_{0}=A_{0}N_{0}$ and on the
exponent $\psi$. \ If $\psi<1/2$, then none of the  $A_{m}$ or $T_{m}$  are
divergent as $h\rightarrow0$. The distance to criticality, parametrized by
$h$, controls the number $m_{\max}$ of angular momentum channels for which
$A_{m}$ and $T_{m}$ are non-negligible ($m_{\max}\sim h^{-1/3}$). On the other
hand, if $\psi>1/2$, then in each  angular momentum channel the amplitude $A$
diverges as $h\rightarrow0$ so that the resulting phase shift saturates at
$\pi/2$ in each channel, and again the number of channels which are relevant
diverges rather strongly as $h\rightarrow0$.

The contribution of this scattering amplitude to the probability that an
electron is scattered through an angle $\theta$,  $W(\theta),$ is given by
$|T(\theta)|^2$.
The impurity scattering rate revealed by the resistivity, 
$\Gamma_{tr}=\int\frac{d\theta}{2\pi}
(1+\cos(\theta))W(\theta)$ (note that we have defined angular coordinates so
that $\theta=0$ corresponds to backscattering). Because we are concerned with
a scattering amplitude strongly peaked about the back-scattering direction,
the $\cos(\theta)$ factor is unimportant. Then $\Gamma_{tr}$ is given by the 
imaginary part of the diagonal $T-$matrix
$\Gamma_{tr} \sim \operatorname{Im}T(0;h)$ in the usual approximation. 
Therefore, we obtain
\begin{equation}
\rho_{res}\sim\operatorname{Im}[T_{0}]+2\sum_{m=1}^{\infty}\operatorname{Im}%
T_{m}\label{rhores}%
\end{equation}
As noted above, within this approximation the resistivity diverges
as criticality is approached, but the
approximation itself breaks down when the mean
free path implied by Eq.\ref{rhores} becomes
smaller than the correlation length. 
The question of a divergent resistivity
is further examined in section VII.

\section{Application to Data}

In this section we attempt to relate calculations of the $h$ dependence of
$\rho_{res}$ to data obtained on $Sr_{3}Ru_{2}O_{7}.$ We first note that it is
sensible to discuss the scattering from an isolated impurity only if the
elastic mean free path $l$ is greater than the correlation length $\xi=\xi
_{0}h^{-1/3}$. The observed low-T resistivity of clean samples of
$Sr_{3}Ru_{2}O_{7}$ is of the order of $5\mu\Omega-cm$ corresponding to a
$p_{F}l\approx250$, i.e. to an $l$ of the order of $300-400\mathring{A}$. The
appropriate value of $\xi_{0}$ is not known at present, but if it is of the
order of the lattice constant or smaller then almost all the available data
are in the regime in which the calculation applies.

\begin{figure}[h]
\includegraphics[height=14cm,width=8cm,angle=0]{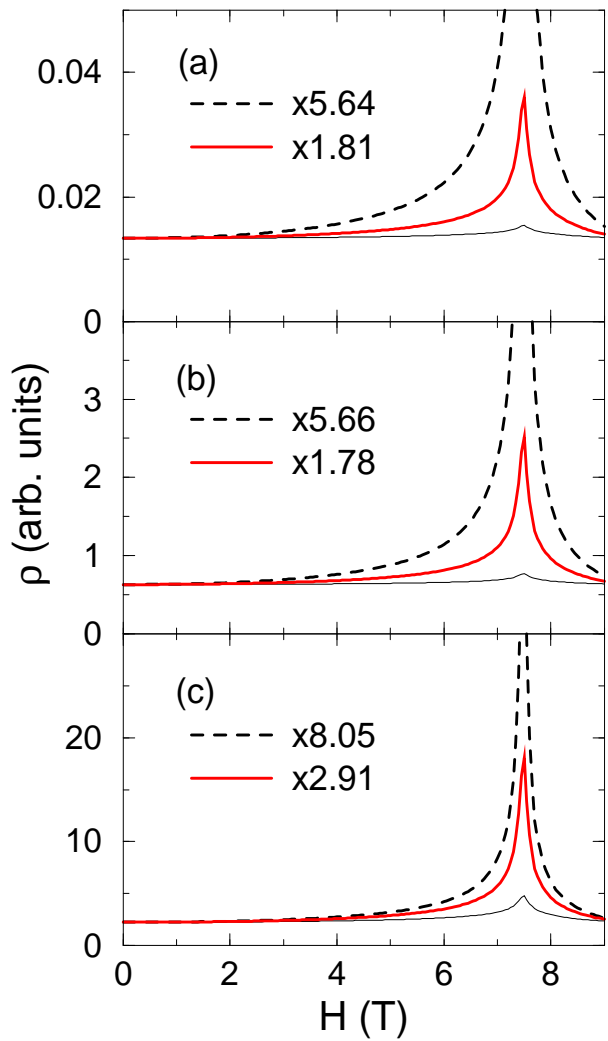}
\caption{Residual resistivity calculated for different exponents and 
initial scattering amplitudes. The curves
for the two larger exponents are scaled (by the factors indicated)
to coincide with the  $H=0$
values calculated for the smallest exponent. 
Dashed lines, $\psi=1$; heavier solid lines,
$\psi=0.75$; light solid lines, $\psi=0.5$. Three panels show the results
of different initial scattering amplitude (a) $g_0=0.5$
(well below unitarity limit for initial point-impurity
scattering amplitude) (b) $g_0=1.0$ (roughly half way
to unitarity limit) (c) $g_0=5.0$ (close to unitarity limit).}
\end{figure}

$Sr_{3}Ru_{2}O_{7}$ presents an interesting issue in modelling. If the
estimates presented in \cite{Millis02a} are correct, then even in zero field
the material is close to a 
ferromagnetic critical point, so that renormalizations of
impurity vertices could be substantial, even at vanishing applied field
(provided, of course, that a sufficiently wide regime exists in which the
$\left\vert {\bf q} \right\vert $ effects of \cite{Belitz98} are not important).
What is unambiguously measurable, however, are 
the effects caused by a varying magnetic field.
Further, the parameters presented in \cite{Millis02a} allow a quantitative
determination of the parameter $h$ above in terms of the applied magnetic
field. We have used the parameters $u_{cc}=-3500K$, $v_{cccc}=58000K$, 
$r=100K$ defined and given below Eq.6 of \cite{Millis02a} to compute the
`mass' in Eq.\ref{ddef} for applied fields from $0$ to $10T$, and have used
this and Eq.\ref{rhores} to calculate the field dependence of the residual
resistivity for various choices of exponent $\psi$ and initial scattering
amplitude. Representative results are shown in Fig.4. The values
at applied field $H=0$ depend on the behavior of the theory at higher
energy scales, which is not known for the reasons given above.
However, the change in 
behavior near the metamagnetic field (about $7T$ for the
parameters used here) should be reliable.
We observe that there is some interplay between exponent
and initial scattering strength but that sharpness and
relative height of the resistivity peak depend most strongly
on the exponent. Comparison to data
\cite{Grigera01} shows that our calculation is consistent with a moderate bare
scattering amplitude and an exponent $\psi\approx3/4$. We also note
that the rather small resistivities observed in experimental systems
suggest either that the initial scattering amplitude is weak
or that the renormalizations associated with the nearby 
ferromagnetic quantum critical point are not large, perhaps
for the reasons given in \cite{Belitz98}.

\section{Friedel Oscillations}

The physics which gives rise to this correction to the conductivity also gives
rise to a singular behavior \cite{Altshuler95} in the susceptibility at
wavevector $q=2p_{F}$ and it is proportional to $(2p_{F}-q)^{1-3\psi}$; this
leads to a change in amplitude and distance dependence of the Friedel
oscillations induced by an impurity in two dimensions. We find that if
$\psi<1/3$ the Friedel oscillations decay as $\cos(2p_{F}R)/R^{\frac{5}%
{2}-{3\psi}}$ while it is $\cos(2p_{F}R)/R^{\frac{1}{2}+{3\psi}}$ when
$\psi>1/3$. Notice that it decays slower (faster) than the standard $1/R^{2}$
form if $1/6<\psi<1/2$ ($\psi<1/6$ or $\psi>1/2$). At non-zero field, the spin
up and spin down Fermi surfaces are characterized by different Fermi wave
vectors so we predict two superposed oscillations emerging from each impurity
site, each decaying with a characteristic power. These oscillations should be
observable in STM experiments. We note, however, that the effect discussed
here is a long wavelength, low energy effect. It does not imply that the
density modulations are 
greatly enhanced near to an impurity; only that they decay
much less (or more) rapidly with distance than in a non-critical material.

It is also interesting to consider the situation at $H=0$, i.e. near to a 
2D ferromagnetic transition, and indeed we note that for the $Sr_{3}Ru_{2}O_{7}$
parameters given in \cite{Millis02a}, $H=0$ corresponds to $h \sim10^{-3}$ so
one might expect the enhancements to be noticeable even at zero field. The
Friedel oscillations from a non-magnetic impurity would indeed be long ranged,
however it is interesting to note that the RKKY interactions are suppressed
(in a system with Heisenberg symmetry) because the spin commutation relations
lead to a minus sign in the renormalization of $\sigma_{z}$ vertex by
$\sigma_{x}$ or $\sigma_{y}$ fluctuations.

We also note that field-dependent STM studies might present an interesting
test of the result \cite{Belitz98} that for 2D Heisenberg materials the
leading momentum dependences is $|\mathbf{q}|$. In this case, the
renormalizations we have discussed would not exist. As $H$ is increased, the
effects that produce $|\mathbf{q}|$ are believed to be cut off, and the
unconventional Friedel oscillations should reappear.

Finally, we observe that the presence of the Friedel oscillations
implies that the state of a critical system is in some sense
a random charge density wave, characterized by infinite ranged
charge oscillations emerging from the various impurity sites.
The resistivity and other properties of such a state are
an interesting issue for future research.

\section{Nonvanishing Impurity Density}

The preceding treatment is valid for correlation lengths less than the mean
free path. We now consider what happens when parameters are tuned so that the
correlation length $\xi$ exceeds the mean free path $l$. We first note that
the problem has two energy scales: the impurity scattering rate $\tau
^{-1}=v_{F}/l$ and the characteristic quantum critical frequency $\omega
^{\ast}=v\xi_{0}^{2}/\xi^{3}$ so that near criticality, when $\xi>\xi_{0}$ we
have $\omega^{\ast}\tau<1$.

For length scales longer than the mean free path the dynamic term in the
action is modified to be \cite{Hertz76}
\begin{equation}
S_{dyn,dirty}=\frac{a^{2}}{2E_{0}} \int{\frac{d^{2}q}{(2\pi)^{2}}T\sum
_{\nu_{n}}\frac{|\nu_{n}|}{D^{\prime}q^{2}}}|\phi(\mathbf{q},i\nu_{n})|^{2}
\label{Sdyndirty}%
\end{equation}
with $D^{\prime}$ of the order of the diffusion constant $D=v_{F}^{2}\tau/2$.
We note $D^{\prime}/D = v/v_{F}$ from the definition of $g^{2}$ in
Eq.\ref{gdef}. Here $1/\tau= 2\pi u^{2}N_0$ is the impurity scattering rate,
$u^{2}$ is the renormalized squared impurity scattering strength (renormalized
due to the effects discussed in section IV), and $N_0$ is the density
of states at the Fermi energy. Thus the dynamical critical exponent is $z=4$
in this case and the nonlinearity in the critical theory remains dangerously irrelevant.

\begin{figure}[h]
\includegraphics[height=2.1cm,width=6cm,angle=0]{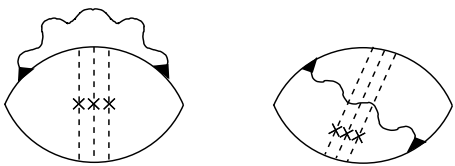}\caption{Diagrams
contributing to the leading order correction to conductivity. The black
triangles and the dotted lines represent the vertex correction by impurity
scattering and the Diffuson ladder.}
\label{figure:5}
\end{figure}

The leading order correction to the conductivity is given by the sum of two
diagrams in Fig.5 \cite{Altshuler80}, where the wavy line represents the
metamagnetic fluctuations with the propagator $D(\mathbf{q},i\nu_{n}%
)=1/(\frac{|\nu_{n}|}{D^{\prime}q^{2}}+q^{2}\xi_{0}^{2}+h^{2/3})$. Notice that
there exist three more diagrams at the same order, but they cancel each other
\cite{Altshuler80}. Each interaction vertex that represents the coupling
between the electrons and the bosonic mode is renormalized by the Diffuson
correction, $(|\nu_{n}|+Dq^{2})^{-1}\tau^{-1}$. 
After evaluating these diagrams we find that the leading
correction to the dc conductivity 
at nonzero  temperature is given by (we assume
$T<1/\tau$)
\begin{align}
\delta\sigma &  =-{\frac{e^{2}}{\hbar}}{\frac{\lambda}{4\pi^{2}}}%
\mathrm{ln}^{2}\left(  {\frac{\lambda D}{T\xi_{0}^{2}}}\right)  \hspace
{1.2cm}\ \mathrm{for}\ T>T_{h}\label{cond1}\\
\delta\sigma &  =-{\frac{e^{2}}{\hbar}}{\frac{\lambda}{3\pi^{2}}}%
\mathrm{ln}\left(  {\frac{1}{h}}\right)  \mathrm{ln}\left(  {\frac{\lambda
D}{T\xi_{0}^{2}}}\right)  \ \ \mathrm{for}\ T<T_{h},\label{cond2}%
\end{align}
where $\lambda=v_{F}/v$ and $T_{h}=h^{4/3}\lambda D/\xi_{0}^{2}$. 
Notice that these results are valid as far as we are in
the perturbative regime, i.e., $\delta \sigma \ll \sigma_0 = 
\frac{e^2}{2 \pi \hbar} k_F l$. 

The physical content of Eqs \ref{cond1},\ref{cond2} is that 
as field is varied at a fixed, low, temperature the resistivity
will initially increase as the critical point is approached,
but sufficiently close to the critical point 
(i.e. when   $h < h_T$ with 
$h_T = (\frac{\xi^2_0 T}{\lambda D})^{3/4}$ )
resistivity will level 
off at a temperature dependent but field-independent value.
Fig. 6 presents a qualitative sketch of the predicted behavior,
including both the 'single-impurity' regime of the previous sections
and the 'Altshuler-Aronov' regime considered in this section.

\begin{figure}[h]
\includegraphics[height=5cm,width=5cm,angle=0]{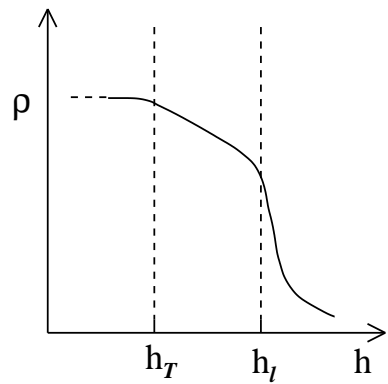}\caption{
Schematic behavior
of  resistivity as a function of $h$ at a fixed temperature. 
Here  $h_l = (\frac{\xi_0}{l})^3$ is the crossover
field at which the mean free path becomes less than
the correlation length and the single-impurity results of section IV
break down, while $h_T = (\frac{\xi^2_0 T}{\lambda D})^{3/4}$ 
is the field scale beyond which the field dependence is cut off
by thermal effects.
If $h_T < h < h_l$, Eq.\ref{cond2} applies
and the resistivity increases logarithmically as $h$ is decreased.
When $h < h_T$, Eq.\ref{cond1} applies and the resistivity becomes a
temperature dependent constant. 
}
\label{figure:6}
\end{figure}

This result should be compared with the well-known behavior of the quantum
correction to the conductivity, $\delta\sigma=-{\frac{e^{2}}{2\pi^{2}\hbar}%
}\mathrm{ln}(1/T\tau)$, in weakly disordered interacting two dimensional
electron systems \cite{Altshuler80}. The notable differences are i)
$\delta\sigma$ in the present case has more singular temperature dependence
when $T>T_{h}$ and at the critical point ($h=0$). ii) If $T<T_{h}$, the
coefficient of the logarithmic term explicitly depends on $h$ and it increases
as the critical point is approached until it is cut off by temperature.

Now let us estimate the size of the perturbative quantum correction in
$Sr_{3}Ru_{2}O_{7}$. We assume that $D^{\prime}\approx D$ and $\xi_{0}$ is of
the order of the in-plane lattice constant $\sim4\mathring{A}$. Using
$v_{F}\sim1eV-\mathring{A}$ and $l\sim300-400\mathring{A}$, we get $1/\tau
\sim30K$ and $D\sim45-60cm^{2}/sec$. Let us consider $h\sim10^{-6}$, then
$T_{h}\sim10^{-3}K$. In this case, $T>T_{h}$ for experimentally relevant
temperature range and Eq.\ref{cond1} should be used. Using the leading order
result, $\sigma_{0}=\frac{e^{2}}{2\pi\hbar}k_{F}l$, in two dimensions the
relative size of the correction can be estimated as $\delta\sigma/\sigma
_{0}\sim0.07$ at $T=2K$ and $\delta\sigma/\sigma_{0}\sim0.1$ at $0.1K$.. Thus
the relative correction is only $7-10\%$. Given that the residual resistivity
is about $5\mu\Omega-cm$, it will be hard to see the effect of these
corrections. If the material were more dirty, say $k_{F}l\sim50$, we would get
$1/\tau\sim150-200K$ and $T_{h}=10^{-4}K$. Similar estimation would predict
that the relative correction is $30-50\%$ at $T=0.1-2K$. It would be
interesting to test these predictions in more dirty samples.

To conclude this section we note that the scattering time $\tau$
is the impurity scattering rate, {\it as renormalized by critical
fluctuations at scales less than the (renormalized) $l$}.
We also observe that our result is perturbative
in both the disorder strength and the interaction. Presumably
when the resistivity becomes of the order of the Mott value
a crossover to insulating behavior occurs. The insulating state
should presumably be interpreted in terms of the random charge
density wave state discussed at the end of Section VI, but
the issue deserves more careful investigation.

\section{Conclusion}

In summary, the effect of critical fluctuations on the residual resistivity is
studied near a two dimensional metamagnetic quantum critical point. When the
correlation length is smaller than the mean free path, the critical
fluctuations induce a singular renormalization of the amplitude of the
back-scattering off an impurity of an electron. This leads to a pronounced
cusp in the residual resistivity near the metamagnetic critical point in
accordance with the experimental results. When the correlation length becomes
larger than the mean free path, the critical fluctuations convert the
`Aronov-Altshuler' logarithmic correction to the conductivity to the more
singular squared-logarithmic behavior near the critical point. Our results
imply a divergent resistivity at criticality. We argued that the
state which gives rise to this divergence is some sort
of random charge density wave, but a detailed investigation
of its properties has not been performed.

Our results may have broader implications. The apparently 
successful comparison of our calculation to data
suggests that the singular ``$2p_F$'' renormalization
discovered in the gauge theory context in \cite{Altshuler94}
is more than a theoretical curiosity, and therefore motivates
examination of other systems where $2p_F$ effects might be
important; for example in the transresistance of bilayer 
$\nu=1/2$ quantum Hall systems. Work in this direction is in
progress.

\textit{Acknowledgements:} We thank A. Mackenzie for stimulating our interest
in this problem and E. Abrahams, P. A. Lee and E. W. Plummer for very helpful
conversations. We are also grateful to L. Taillefer, F. Ronning, and R. Hill
for showing their unpublished data and helpful discussions. This work was
supported by NSF-DMR-00081075 (AJM), the Alfred P. Sloan Foundation and the
NSERC of Canada (YBK), and the Canadian Institute for Advanced Research. We
acknowledge the hospitality of the Aspen Center for Physics, where part of
this work was performed.

\end{document}